# Magnetic properties of Hydrogenated Li and Co doped ZnO nanoparticles

O. D. Jayakumar, I. K. Gopalakrishnan*, K. Shasikala, and S. K. Kulshreshtha

Chemistry Division, Bhabha Atomic Research Centre, Mumbai 400085, India.

C. Sudakar[#]

[#]Department of Physics and Astronomy, Wayne State University, Detroit, MI 48201,U.S.A.

## Abstract

The effect of hydrogenation on magnetic properties of $Zn_{0.85}Co_{0.05}Li_{0.10}O$ nanoparticles is presented. It was found that the sample hydrided at room temperature (RT) showed weak ferromagnetism (FM) while that hydrided at $400^oC$ showed robust ferromagnetism at room temperature. In both cases reheating the sample at $400^oC$ in air converts it back into paramagnetic state (P) completely. The characterization of samples by X-ray and electron diffraction (ED) showed that room temperature ferromagnetism observed in the samples hydrogenated at RT is intrinsic in nature whereas that observed in the samples hydrogenated at $400^oC$ is partly due to the cobalt metal clusters.

Corresponding author Email: ikopal@barc.gov.in

ZnO is a wide band gap semiconductor with optoelectronic properties that make it an attractive candidate for a variety of device applications. Besides, the dilute magnetic semiconductor (DMS) effect observed in transition metal (TM) doped ZnO has triggered a flurry of activity[1-3]. It is well documented that for the appearance of ferromagnetism in TM doped ZnO bulk and nanocrystalline samples, the presence of additional carries, in addition to the doped magnetic ions is a must[4-6]. This can be generated either by additional carrier doping or by generation of defects. ZnO is an n-type semiconductor. N-type conductivity has been attributed to the presence of interstitial Zn and the hydrogen contamination during its growth. Hydrogen is present in many of the techniques commonly used for growth of ZnO, such as hydrothermal growth[7], metal-organic chemical vapor deposition (MOCVD)[8] or vapor phase transport[9]. When the growth is carried out in air, water vapor acts as a source of hydrogen. Techniques such as laser ablation or sputtering are sometimes intentionally carried out in a hydrogen atmosphere. In addition, water vapor is always present as a residual gas in any high vacuum system, serving as a source of hydrogen in techniques such as molecular beam epitaxy. It is also possible that hydrogen can penetrate the crystal when exposed to water. Recent theoretical calculations indicate that isolated H always behaves as a shallow donor in ZnO, leading to the suggestion that H is the cause of the unintentional $n$-type conductivity[10]. This work has catalyzed the interest in H in ZnO and spurred a variety of recent studies that confirm hydrogen's role as a shallow donor[11-15]. From EPR and ENDOR studies at 9.5 GHz by Hofmann et. al.[11] show that one of the two observed donor resonances in ZnO single crystals is related to hydrogen. The H nucleus is in the center or in a position close to the center of the shallow donor electron wave function.

The observed hyperfine interactions with about 50 shells of Zn nuclei prove that it is a shallow, effective-mass-like donor with a Bohr radius of about 1.5 nm. Recently it was shown that ferromagnetism can be activated in TM doped ZnO and $TiO_2$ by subjecting it to treatment in hydrogen environment[16-19]. Recently[20], we have shown that high-$T_c$ ferromagnetism can be activated in $Zn_{0.85}Co_{0.05}Li_{0.10}O$ nanocrystalline samples by surfactant treatment. In this communication, we show the reversible cycling of paramagnetism and ferromagnetism on RT and 400°C hydrogenated samples of $Zn_{0.85}Co_{0.05}Li_{0.10}O$ by thermal treatments.

The samples of nominal composition $Zn_{0.85}Co_{0.05}Li_{0.10}O$ were synthesized by pyrolysis of mixture of zinc acetate di hydrate, cobalt acetate tetra hydrate, lithium acetate and sodium bi carbonate taken in appropriate proportions at 175°C for 2 hours, the details of which are given elsewhere[20,21]. For hydrogen treatment, samples were loaded in a stainless steel Sievert type reactor of hydriding and dehydriding unit. The unit is evacuated to $10^{-6}$ torr. For RT hydriding the sample was heated to 400°C for 2 hrs under vacuum and subsequently cooled to room temperature followed by applying a hydrogen pressure of around 5-10 atm overnight. Then the sample is taken out of reactor after cooling it to Liquid $N_2$ temperature. The excess $H_2$ was then removed by evacuation followed by passivating the sample by exposing it to air at liquid $N_2$ temperature. The temperature was then to raised to RT before taking out the sample to atmosphere. For hydriding at 400°C, same procedure is followed except that the hydriding is carried out at 400°C instead of RT. Phase purity and the structure of the samples were analysed using CuKα radiation by employing a Philips Diffractometer (model PW 1071). The lattice parameters of the compounds were extracted by Rietveld refinement of the XRD data by using the program Fullprof [22]. High-resolution electron microscopy (HRTEM) imaging, selected area electron diffraction (SAED) and energy-dispersive X-ray spectroscopy (EDS) studies were carried out with a JEOL, JEM 2010 transmission electron microscope (TEM). *DC* magnetization measurements were carried out using an E.G.&G P.A.R vibrating sample magnetometer (model 4500).

The XRD patterns of as synthesized $Zn_{0.85}Co_{0.05}Li_{0.10}O$ and as well as after hydrogenation and de-hydrogenation cycles showed they are single phase without any stray peaks corresponding to parasitic phases (graph 1-supplementary material). It is worth noting that the crystallite size decreases after first and second cycle of hydriding. The lattice parameter values obtained by Rietveld refinement (graph 2- supplementary material) of as synthesized $Zn_{0.85}Co_{0.05}Li_{0.10}O$ are a = 3.2630(2) and b =5.2179(1) Å. After second cycle of hydriding at 400$^o$C the lattice parameter values decreased to a = 3.2380(2) and b = 5.2140(1) Å. This may be due to the reduction of CoO into Co metal nanoparticles which is in agreement with the ED studies discussed in the following paragraph. However, it is worth noting that the lattice parameter of the sample hydrogenated at RT (a = 3.2651(2) , b = 5.2180(1)) did not change much in comparison that of as synthesized sample.

Three different samples of Co, Li doped ZnO (as synthesized, $H_2$ treated at RT and $H_2$ treated at 400$^o$C)) were characterized by TEM, HRTEM, SAED and EDS analyses. As synthesized samples were highly agglomerated with the monocrystalline particles of 5-50 nm in size (Fig.1a). The SAED pattern (Fig. 1c) from a finely dispersed region shows the rings made up of discrete spots indicating the highly crystalline nature of the nanoparticulates. The SAED pattern is consistent with the wurtzite ZnO structure with intense ring patterns from (hk*i*l) planes as indexed in the electron diffraction pattern. No rings corresponding to secondary phases were evidenced from SAED pattern. The HRTEM images of these particles show defect free structure (Fig.1b) and the composition determined by EDS (Fig.1d) was uniform for these samples. The samples subjected to hydrogenation at RT also showed similar results. In contrast, the samples treated with $H_2$ at 400$^o$C though retained the morphological features (Fig.2a), the particles had developed nano-pores within the crystallites (graph 3-supplementary material) and with large defect structures (Fig.2b). SAED pattern (Fig.2c) reveals the presence of Co nanoparticles in the sample.

Figure 3 shows the field dependence of magnetization measured at room temperature for $Zn_{0.85}Co_{0.05}Li_{0.10}O$ hydrided at room temperature and 400$^o$C. In addition *M-H* loops of as prepared and undoped ZnO is also given. It can be seen that for the sample $Zn_{0.85}Co_{0.05}Li_{0.10}O$ without hydrogen treatment the magnetization is linear indicating its paramagnetic nature, while that treated with hydrogen are all *S* shaped typical of ferromagnetic materials. The saturation magnetization of sample treated with hydrogen at room temperature ($M_s$) is 0.058 emu/g (0.0191$\mu_B$/Co). It can be seen that there is hardly any change in magnetic moment on repeated hydriding. However, on raising the temperature of hydriding to 400$^o$C the samples became strongly ferromagnetic. The $M_s$ value of sample increases from 0.058 emu/g to 0.19 emu/g (after first cycle) on raising the temperature from RT to 400$^o$C. On dehydriding the sample by heating in air it became paramagnetic (Fig.3, inset 'h'). On further hydriding this paramagnetic sample became ferromagnetic with enhanced $M_s$ value (Fig.3, inset 'g').

XRD and electron diffraction (ED) data of $Zn_{0.85}Co_{0.05}Li_{0.10}O$ before and after hydrogenation at room temperature showed it is single phase without any secondary phase inclusions. In view of this the weak ferromagnetism observed in the sample can be attributed to the incorporation of hydrogen into the ZnO lattice. Park and Chadi[23] using first principle pseudopotential calculations predicted that H can induce ferromagnetic spin-spin interaction between neighbouring magnetic impurities (TM) through the formation of a bridge bond. However, this presumption may not be valid for the sample hydrided at 400°C in view of the presence of Co metal as secondary phase as evidenced by the SAED shown in Fig2c. The observation of nano-pores in the HRTEM (Fig.2b) also points to the reduction of CoO into Co by the incorporation of hydrogen. Recently bulk samples and thin films of Co doped ZnO hydrided at elevated temperature showed the presence of metal clusters[17-19]. In addition many authors have found the presence of Co metal clusters in thin films of Co doped ZnO made under high vacuum conditions.

It is well documented that the hydrogenation of ZnO at RT generates donors which decay with time[24,25]. However, if the hydrogenation is carried out at elevated temperatures stable carriers are generated[26]. In view of this it is tempting to attribute the robust ferromagnetism observed in samples hydrogenated at 400°C to the incorporation of hydrogen into ZnO lattice. However, in view of the presence of Co metal clusters in the sample hydrogenated at 400°C as evidenced by ED one cannot rule out the origin of ferromagnetism from Co metal clusters. It is pertinent to note that saturation $M_s$ value of hydrogenated sample is much less than that is expected of Co metal (1.7$\mu_B$). In addition in a very recent report, Lee et al.[18] did not detect ferromagnetic contribution from Co metal from the magnetic circular dichroism studies, although Co metallic peak was observed in X-ray photoelectron spectroscopy spectra of the $H_2$ reduced samples. Lee *et al.* showed that Co 9.1% ZnO and Co 5% ZnO thin films become ferromagnetic when hydrogenated with Ar–$H_2$ mixed gas, due to the enhanced ferromagnetic spin-spin interaction in the H–Co coupling. Deka and Joy[19] have observed RTF in polycrystalline $Zn_{0.95}Co_{0.05}O$ hydrogenated at 1125 K with $M_s$ value as high as 9.5 emu/g. This comes to ~2.7 $\mu_B$/Co. This value is very much higher than that is expected for metallic Co (1.7$\mu_B$/Co). However, this value is close to the ideal value of 3$\mu_B$ for $Co^{2+}$-ions in a tetrahedral crystal field. Their optical studies also showed that hydrogenation did not affect the substituted $Co^{2+}$ ions inside the wurtzite crystal lattice. In view of above factors one can attribute the enhanced RTF observed in $Zn_{0.85}Co_{0.05}Li_{0.10}O$ hydrogenated at 400°C as arising partly from the Co metal clusters and partly due to the incorporation of carriers subsequent upon incorporation of hydrogen into doped ZnO lattice. It is worth noting that Sung et al.[27] observed reproducible giant magnetic moment of 6.1 $\mu_B$/$Co^{2+}$ ion, at room temperature in $Zn_{0.96}Co_{0.04}O$ films grown by reactive magnetron co-sputtering on $LiNbO_3$ substrates under vacuum ($5\times10^{-6}$ torr).

In summary, we have investigated the effect of hydriding on the magnetism of Co-Li-Zn-O system. It was observed that hydriding the samples at RT, induce weak ferromagnetism whereas that hydrided at 400°C showed strong ferromagnetic bahavior. XRD and ED analyses of the samples showed that they are single phase when hydrided at RT. However on raising the temperature of hydriding to 400°C Co metal clusters start appearing in the samples as evidenced by ED analyses.

# References

1. P. Sharma, A. Gupta, K. V. Rao, F. J. Owens, R. Sharma, R. Ahuja, J. M. O. Guillen, B. Johansson and G. A. Gehring, Nature Materials, **2**, 673 (2003).

2. K. R. Kittilstved, W. K. Liu, and D. R. Gamelin, Nature Materials, **5**, 291 (2006).

3. K. R. Kittilstved, D.A. Schwartz, A.C. Tuan, S. M. Heald, S. A. Chambers and D. R. Gamelin, Phys. Rev. Lett., **97**, 037203 (2006).

4. S. Ramachandrana, J. Narayan and J. T. Prater, Appl. Phys. Lett. **88**, 242503 (2006).

5. X. Wang, J.B. Xu, N. Ke, J. Yu, J. Wang, Q. Li, H.C. Ong and R. Zhang, Appl. Phys. Lett., **88**, 223108 (2006).

6. J. R. Neal, A. J. Behan, R. M. Ibrahim, H. J. Blythe, M. Ziese, A. M. Fox and G. A Gehring, Phys. Rev. Lett. **96**, 197208 (2006).

7. W. J. Li, E. W. Shi, W. Z. Zhong and W. Z. Tin, J. Cryst. Growth, **203**, 186 (1999).

8. B. P. Zang, N. T. Binh, K. Wakatsuki, Y. Segawa, Y. Kashiwaba and K. Haga, Nanotechnogy **15**, S382 (2004).

9  J. Q. Hu, Q. Li, N. B. Wong, C. S. Lee and S. T. Lee, Chem. Mater. **14**, 1216 (2002).

10. S.F.J. Cox, E.A. Davis, P.J.C King, J.M. Gil, H.V. Alberto and R.C. Vilao. J Phys: Cond Mater, **13**, 9001 (2001).

11. D.M. Hofmann, A. Hofstaetter, F. Leiter, H. Zhou, F. Henecker and B.K. Meyer, Phys Rev Lett., **88**, 045504 (2002).

12. S. F. J. Cox, E. A. Davis, S. P. Cottrell, P. J. C. King, J. S. Lord, J. M. Gil, H. V. Alberto, R. C. Vilão, J. Pironto Duarte, N. Ayres de Campos, A. Weidinger, R. L. Lichti, and S. J. C. Irvine, Phys. Rev. Lett. **86**, 2601 (2001).

13. H. Sheng, S. Muthukumar, N.W. Emanetoglu and Y. Lu. Appl Phys Lett., **80**, 2132 (2002).

14. K. Ip, M. E. Overberg, Y.W. Heo, D.P. Norton, S.J. Pearton and C.E. Stutz, Appl. Phys Lett., **82,** 385 (2003).

15. F. D. Auret, S.A. Goodman, M. Hayes, M.J. Legodi, H.A. Van Laarhoven and D.C. Look. Appl. Phys Lett., **79**, 3074 (2001).

16. A.Manivannan, G. Glospel, P. Dutta and M. S. Seehra Apply. Phys. Lett. **97**, 10D325 (2005).

17. Y. Wang. , L. Sun, L-G. Kong, J-F. Kang, X. Zhang and R-Q. Han Journal of Alloys and Compounds (2006) in print.

18. H.J. Lee, C.H. Park, S.-Y. Jeong, K.-J. Yee, C.R. Cho, M.-H. Jung and D.J. Chadi, Apply. Phys. Lett., **88**, 062504 (2006).

19. S. Deka and P.A. Joy, Apply. Phys. Lett., **89**, 32508 (2006).

20. O. D. Jayakumar, I. K. Gopalakrishnan and S. K. Kulshreshtha Adv. Mater.**18**, 1867 (2006).

21. Z. Wang, H. Zhang, L. Zhang, J. Yuan, S. Yan and C. Wang, Nanotechnology, **14**, 11 (2003).

22. J. Rodriguez-Carvajal, *Fullprof: a program for Rietveld Refinement and Profile Matching analysis of Complex Powder Diffraction Patterns ILL.* XV Congress of the IUCr, International Union of Crystallography, Toulouse, France, **1990**, p. 127.

23. H. Park and D. J. Chadi, Phys. Rev. Lett. **94**, 127204 (2005).

24. McCluskey, S.J Jokela. and W.M.H. Oo,. Physica B **376-377**, 690 (2006).

25. M. G. Wardle, J. P. Goss, and P. R. Briddon, Phys Rev Lett. **96**, 205504 (2006).

26. G. A Shi Ml Stavola, S. J. Pearton, M. Thieme, E. V. Lavrov and J. Weber, Phys. Rev. B **72**, 195211 (2005).

27  C. Song, K. W. Geng, F. Zeng, X. B. Wang, Y. X. Shen, F. Pan, Y. N. Xie, T. Liu, H. T. Zhou, and Z. Fan, Phys. Rev. B **73**, 024405 (2006).

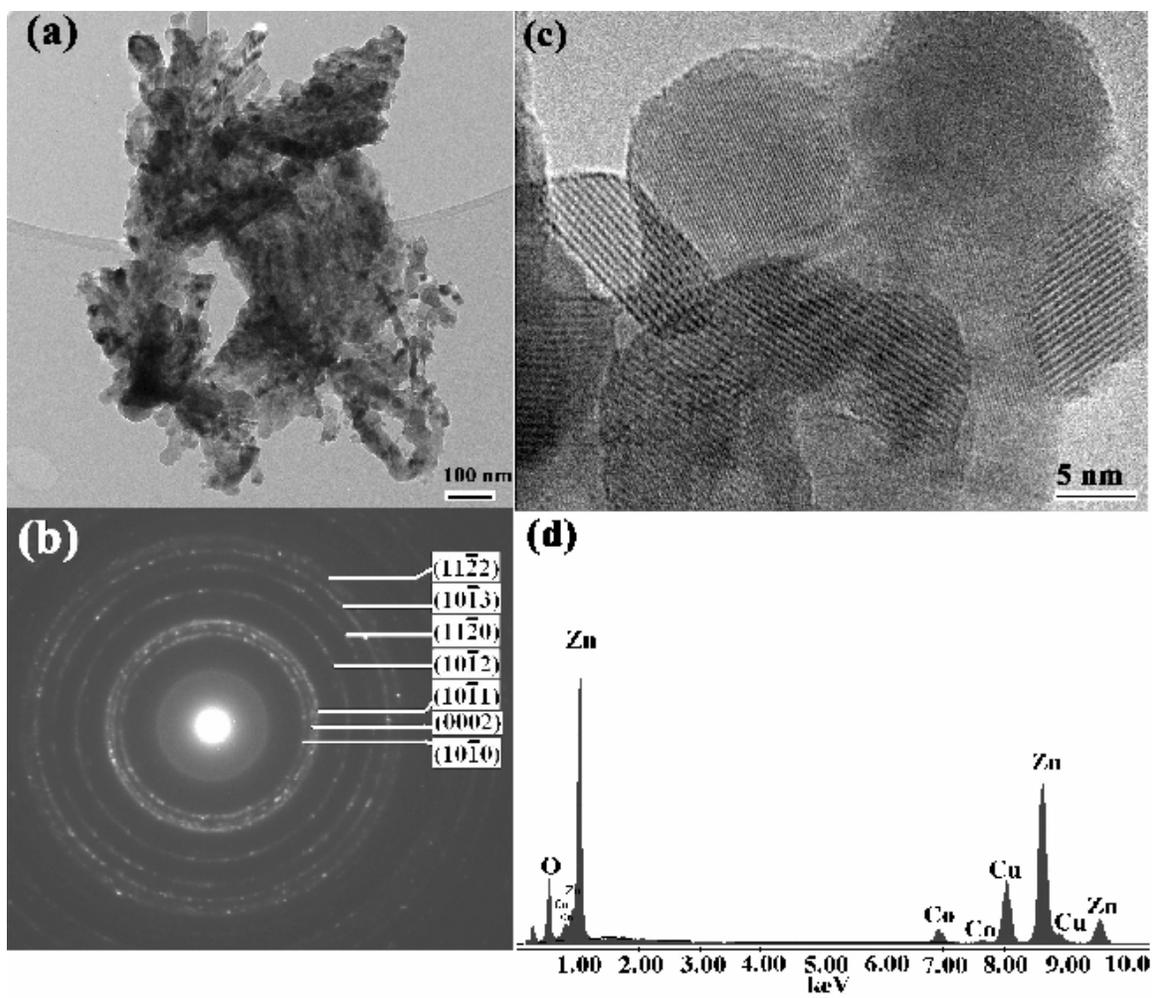

Figure.1 a)TEM image, b) HRTEM image, c)SAED and d) EDS of as synthesized $Zn_{0.85}Co_{0.05}Li_{0.10}O$.

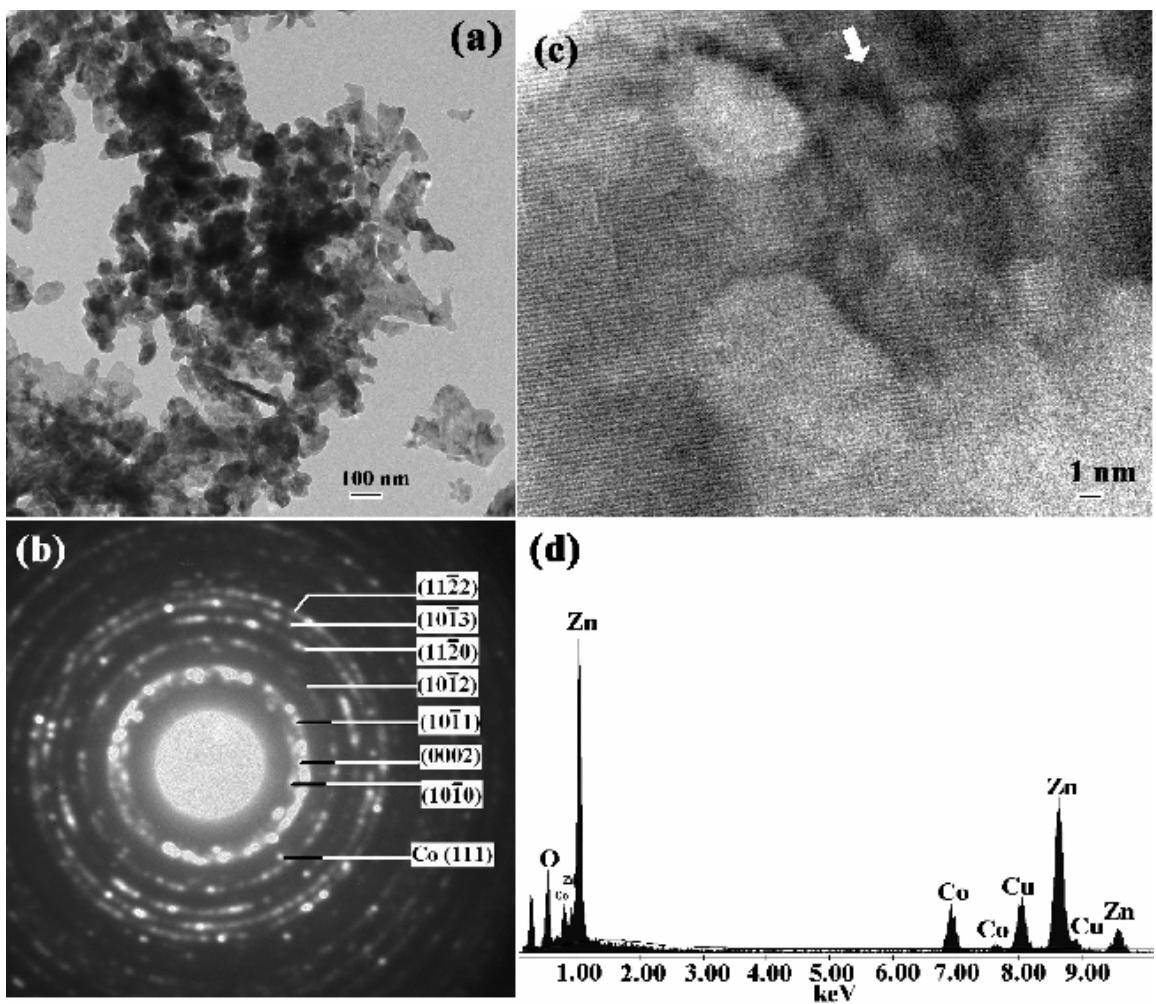

Figure.2 a)TEM image, b) HRTEM image (dark patches marked by arrow indicate regions of Co metal clusters), c) SAED and d) EDS of $Zn_{0.85}Co_{0.05}Li_{0.10}O$ after second cycle of hydrogenation at $400^oC$.

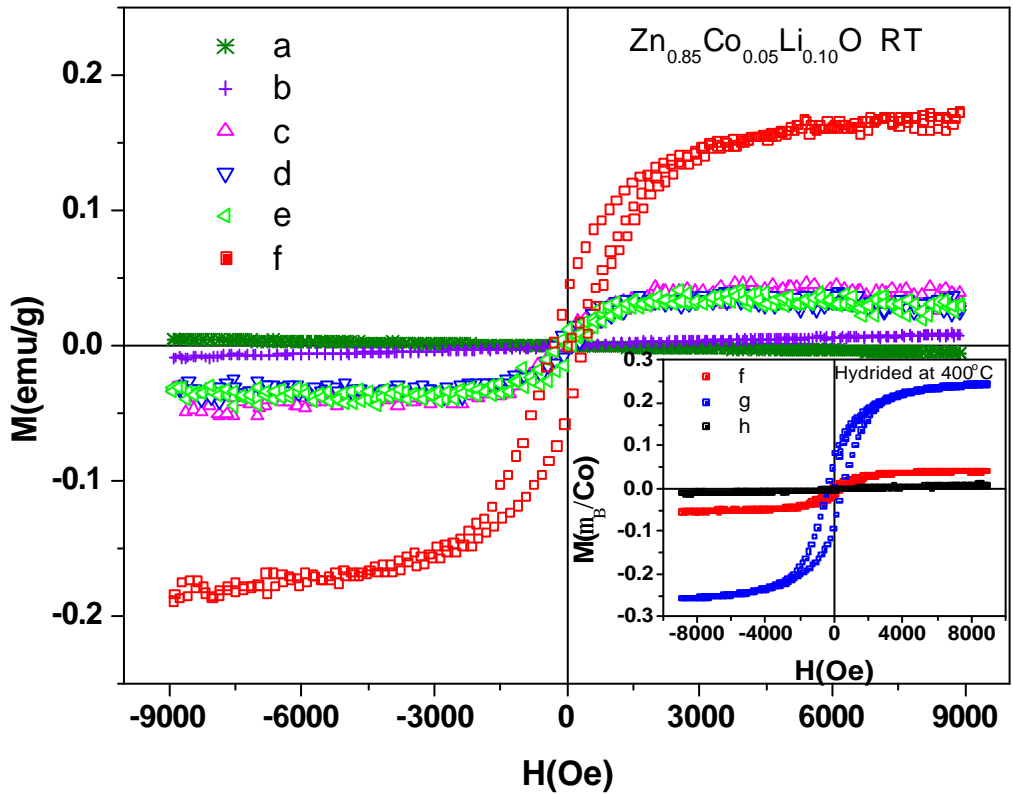

Fig. 3 M vs H curves at RT for a) Pristine ZnO b) as synthesized $Zn_{0.85}Co_{0.05}Li_{0.10}O$ c) after first cycle of hydriding at RT of $Zn_{0.85}Co_{0.05}Li_{0.10}O$ d) after second cycle of hydriding at RT of $Zn_{0.85}Co_{0.05}Li_{0.10}O$ e) after third cycle of hydriding at RT of $Zn_{0.85}Co_{0.05}Li_{0.10}O$ f) after first cycle of hydriding at $400^{o}C$ of $Zn_{0.85}Co_{0.05}Li_{0.10}O$. The inset shows M-H loops of f) after first cycle of hydriding at $400^{o}C$ of $Zn_{0.85}Co_{0.05}Li_{0.10}O$ g) after second cycle of hydriding at $400^{o}C$ of $Zn_{0.85}Co_{0.05}Li_{0.10}O$ and h) after heating $Zn_{0.85}Co_{0.05}Li_{0.10}O$ at $400^{o}C$ in air after first cycle of hydriding.